\documentclass[fleqn,usenatbib]{mnras}

\usepackage{newtxtext,newtxmath}

\usepackage[T1]{fontenc}

\DeclareRobustCommand{\VAN}[3]{#2}
\let\VANthebibliography\thebibliography
\def\thebibliography{\DeclareRobustCommand{\VAN}[3]{##3}\VANthebibliography}

\usepackage{graphicx}	
\usepackage{amsmath}	
\usepackage{xcolor}
\usepackage{xspace}
\usepackage{ulem}
\usepackage{comment}
\usepackage{anyfontsize} 
\usepackage{lineno}
\usepackage{orcidlink}

\newcommand{\XL}{{\it XL-Calibur}\xspace}
\newcommand{\I}{{\it IXPE}\xspace}
\newcommand{\pogo}{{\it PoGO+}\xspace}

\newcommand{\Nu}{{\it NuSTAR}\xspace}
\newcommand{\C}{{\it Chandra}\xspace}
\newcommand{\Hit}{{\it Hitomi}\xspace}

\title[XL-Calibur hard X-ray polarimetry of the Crab]{\textit{XL-Calibur} measurements of polarised hard X-ray emission from the Crab}

\author[H. Awaki et al.]{
Hisamitsu Awaki,$^{1}$
Matthew G. Baring\orcidlink{0000-0003-4433-1365},$^{2}$
Richard Bose,$^{3}$
Dana Braun,$^{3}$
Jacob Casey\orcidlink{0009-0009-3051-6570},$^{4}$
\newauthor
Sohee Chun\orcidlink{0009-0002-2488-5272},$^{3}$
Pavel Galchenko\orcidlink{0000-0001-6358-5147},$^{5}$
Ephraim Gau\orcidlink{0000-0002-5250-2710},$^{3}$\thanks{E-mail: ephraimgau@wustl.edu}
Kazuho Goya,$^{6}$
Tomohiro Hakamata\orcidlink{0000-0002-0987-0278},$^{7}$
\newauthor
Takayuki Hayashi,$^{8,9}$
Scott Heatwole,$^{5}$
Kun Hu\orcidlink{0000-0002-9705-7948},$^{3}$
Ryo Imazawa\orcidlink{0000-0002-0643-7946},$^{6}$
Daiki Ishi,$^{10}$
Manabu Ishida,$^{10}$
\newauthor
Fabian Kislat\orcidlink{0000-0001-7477-0380},$^{4}$
M\'ozsi Kiss\orcidlink{0000-0001-5191-9306},$^{11,12}$\thanks{E-mail: mozsi@kth.se}
Kassi Klepper,$^{11,12}$
Henric Krawczynski\orcidlink{0000-0002-1084-6507},$^{3}$
Haruki Kuramoto,$^{7}$
\newauthor
R. James Lanzi,$^{5,8}$
Lindsey Lisalda\orcidlink{0000-0002-5202-1642},$^{3}$
Yoshitomo Maeda\orcidlink{0000-0002-9099-5755},$^{10}$
Filip af Malmborg\orcidlink{0009-0007-2274-2055},$^{11,12}$
\newauthor
Hironori Matsumoto,$^{7,13}$
Shravan Vengalil Menon\orcidlink{0009-0005-0818-7484},$^{3}$
Aiko Miyamoto,$^{7}$
Asca Miyamoto\orcidlink{0009-0008-4216-064X},$^{14}$
\newauthor
Takuya Miyazawa\orcidlink{0000-0002-6068-6337},$^{15}$
Kaito Murakami,$^{7}$
Azuki Nagao,$^{7}$
Takashi Okajima\orcidlink{0000-0002-6054-3432},$^{8}$
Mark Pearce\orcidlink{0000-0001-7011-7229},$^{11,12}$\thanks{E-mail: pearce@kth.se}
\newauthor
Brian F. Rauch\orcidlink{0000-0002-1452-4142},$^{3}$
Nicole Rodriguez Cavero\orcidlink{0000-0001-5256-0278},$^{3}$
Kohei Shima,$^{7}$
Kentaro Shirahama,$^{7}$
Carlton M. Snow,$^{5}$
\newauthor
Sean Spooner\orcidlink{0000-0003-0710-8893},$^{4}$
Hiromitsu Takahashi\orcidlink{0000-0001-6314-5897},$^{6}$
Sayana Takatsuka,$^{7}$
Keisuke Tamura,$^{8,9}$
Kojiro Tanaka,$^{14}$
\newauthor
Yuusuke Uchida\orcidlink{0000-0002-7962-4136},$^{16}$
Andrew Thomas West\orcidlink{0000-0002-5471-4709},$^{17}$
Eric A. Wulf\orcidlink{0000-0002-9577-7888},$^{18}$
Masato Yokota,$^{6}$
and Marina Yoshimoto\orcidlink{0009-0005-0819-0819}$^{7}$
\\
$^{1}$Graduate School of Science and Engineering, 2-5, Bunkyo-cho, Matsuyama, Ehime 790-8577, Japan\\
$^{2}$Department of Physics and Astronomy -- MS 108, Rice University, 6100 Main Street, Houston, Texas 77251-1892, USA\\
$^{3}$Department of Physics, McDonnell Center for the Space Sciences and Center for Quantum Sensors, Washington University in St. Louis, 1 Brookings Dr, \\Saint Louis, MO 63130, USA\\
$^{4}$Department of Physics and Astronomy, and Space Science Center, University of New Hampshire, 8 College Rd, Durham, NH 03824, USA\\
$^{5}$NASA Wallops Flight Facility, Fulton St, Wallops Island, VA 23337, USA\\
$^{6}$Graduate School of Advanced Science and Engineering, Hiroshima University, 1-3-1 Kagamiyama, Higashi-Hiroshima, Hiroshima 739-8526, Japan\\
$^{7}$Department of Earth and Space Science, Osaka University, 1-1 Machikaneyama-cho, Toyonaka, Osaka 560-0043, Japan\\
$^{8}$NASA Goddard Space Flight Center, Greenbelt, MD 20771, USA\\
$^{9}$University of Maryland, Baltimore County, 1000 Hilltop Circle, Baltimore, MD 21250, USA\\
$^{10}$Japan Aerospace Exploration Agency, Institute of Space and Astronautical Science, 3-1-1 Yoshino-dai, Chuo-ku, Sagamihara, Kanagawa 252-5210, Japan\\
$^{11}$KTH Royal Institute of Technology, Department of Physics, 106 91 Stockholm, Sweden\\
$^{12}$The Oskar Klein Centre for Cosmoparticle Physics, AlbaNova University Center, 106 91 Stockholm, Sweden\\
$^{13}$Forefront Research Center, Graduate School of Science, Osaka University, Japan\\
$^{14}$Department of Physics, Tokyo Metropolitan University, 1-1 Minami-Osawa, Hachioji, Tokyo 192-0397, Japan\\
$^{15}$Okinawa Institute of Science and Technology Graduate University, 1919-1 Tancha, Onna-son, Kunigami-gun Okinawa, 904-0495 Japan\\
$^{16}$Tokyo University of Science, 2641 Yamazaki, Noda, Chiba 278-8510, Japan\\
$^{17}$University of Arizona, Department of Astronomy, Steward Observatory, 933 North Cherry Avenue, Tucson, AZ 85721-0065, USA\\
$^{18}$U.S. Naval Research Laboratory, 4555 Overlook Avenue, SW Washington, DC 20375, USA
}

\date{Accepted XXX. Received YYY; in original form ZZZ}

\pubyear{\the\year{}}

\begin{document}
\label{firstpage}
\pagerange{\pageref{firstpage}--\pageref{lastpage}}
\maketitle

\begin{abstract}
We report measurements of the linear polarisation degree (PD) and angle (PA) for hard X-ray emission from the Crab pulsar and wind nebula. Measurements were made with the \XL ($\sim$15--80~keV) balloon-borne Compton-scattering polarimeter in July 2024. The polarisation parameters are determined using a Bayesian analysis of Stokes parameters obtained from X-ray scattering angles. 
Well-constrained ($\sim$8.5$\sigma$) results are obtained for the polarisation of the $\sim$19--64~keV signal integrated over all pulsar phases: PD=(25.1$\pm$2.9)\% and PA=(129.8$\pm$3.2)$^\circ$. 
In the off-pulse (nebula-dominated) phase range, the PD is constrained at $\sim$4.5$\sigma$ and is compatible with the phase-integrated result. 
The PA of the nebular hard X-ray emission aligns with that measured by \I in the 2--8~keV band for
the toroidal inner region of the pulsar wind nebula, where the hard X-rays predominantly originate.
For the main pulsar peak, PD=(32.8$^{+18.2}_{-28.5}$)\% and PA=(156.0 $\pm$ 21.7)$^\circ$, while for the second peak (inter-pulse), PD=(0.0$^{+33.6}_{-0.0}$)\% and PA=(154.5 $\pm$ 34.5)$^\circ$. 
A low level of polarisation in the pulsar peaks likely does not favour emission originating from the inner regions of the pulsar magnetosphere.
Discriminating between Crab pulsar emission models will require deeper observations, e.g. with a satellite-borne hard X-ray polarimeter. 
\end{abstract}

\section{Introduction}
The Crab, comprising a rotation-powered pulsar and wind nebula (PWN), is one of the brightest persistent sources of celestial X-rays~\citep{buhler2014surprising}.
The pulsar is a highly magnetised ($\sim$10$^{12}$~G) neutron star with a rotation period $\sim$33.8~ms. 
A small fraction of the rotational energy powers a relativistic leptonic wind, which illuminates the nebulous supernova remnant yielding synchrotron and inverse Compton emission across the electromagnetic spectrum~\citep{Gaensler.2006,Hester.2008}. 
The Crab has been extensively studied using spectroscopy, timing and imaging, but the addition of X-ray polarimetry allows the emission to be studied in a systematically new way.
The linear polarisation of the emission is described by a polarisation degree (PD, \%) and a polarisation angle (PA, $^{\circ}$)\footnote{All source-related angles reported in our paper are defined relative to celestial north, going anticlockwise (i.e. to the East).}. In the soft X-ray band, 2--8~keV, the linear polarisation of the X-ray emission was recently measured by the \I satellite mission~\citep{Bucciantini.2023,Wong_2023,MizunoIXPE,wong2024analysiscrabxraypolarization,ixpeopticalandxray}.
Here, we present complementary measurements in the neighbouring hard X-ray band, {$\sim$19--64~keV}, from the balloon-borne mission \XL. 

The pulsar X-ray light-curve has two peaks, as is the case across the electromagnetic spectrum. The main pulse (P1) is separated from a second pulse (P2, also referred to as the inter-pulse) by a bridge region. The remaining off-pulse (OP) region is dominated by nebula emission. 
A fundamental understanding of the emission locale and field geometry for the pulsar is lacking~\citep{Harding2019book}.
The evolution of PD and PA across the light curve elucidates the geometry of regions where accelerated particles dissipate their energy, e.g. potentially discriminating between emission from just inside the light cylinder and proximate to the magnetospheric current sheet~\citep{Cerutti.2016pz1,Harding.2017feh}, and that outside the magnetosphere in the inner nebula~\citep{Petri2013,Harding.2017feh}.

Spectroscopic and spectro-polarimetric images of the Crab constrain the structure of the PWN, the process of particle acceleration at relativistic shocks, and the propagation of particles in the downstream plasma.
\C images~\citep{Weisskopf.2000} reveal complex structure in the inner nebula, including time-varying polar jets, an equatorial torus and shock structures. Two concentric magnetic tori are centred on the pulsar. The inner torus lies in a plane perpendicular to the pulsar spin-axis, which has a sky-projected angle of (124.0$\pm$0.1)$^{\circ}$~\citep{Ng_2004}.

In a phase-resolved analysis of \I data (300 ks), \cite{wong2024analysiscrabxraypolarization} revealed a +40$^\circ$ PA swing throughout P1, where the maximum PD$\sim$15\% is reached close to the peak phase. 
The phase-dependence of polarisation parameters was found to differ significantly from the optical band, indicating that different mechanisms or locations are responsible for the polarised emission in the two bands. 
In contrast,~\citet{ixpeopticalandxray} present a phenomenological model connecting the polarisation properties in the soft X-ray and optical bands. This suggests a common underlying emission mechanism for both bands. 

In ~\citet{Bucciantini.2023} (92~ks), \I reports PD=(19.0$\pm$0.2)\% and PA=(145.5$\pm$0.3)$^\circ$ for phase-integrated emission measured within 2.5~arcmin of the pulsar. Similar off-pulse values are found for the region.
The polarisation map follows expectations for synchrotron emission in the 100-150~$\mu$G toroidal magnetic field estimated in~\citet{MizunoIXPE}. 
Local enhancement of the PD in several regions (up to 40--50\% to the north and south of the inner nebula) is attributed to variations in the amount of magnetic turbulence within the nebula. 
For the nebula-dominated off-pulse region measured within 20 arcsec of the pulsar, \citet{Bucciantini.2023} reports \mbox{PD=(24.1$\pm$0.8)\%} and \mbox{PA=(133.6$\pm$1.0)$^\circ$}. 
In the deeper \I observations reported in~\citet{wong2024analysiscrabxraypolarization}, the polarisation map of the inner nebula confirms the presence of a toroidal magnetic field. 
A spatial correlation is found between polarisation parameters and photon spectral index, which informs the interpretation of \XL observations presented here, where the magnetic field orientations are sampled by higher-energy electrons in the inner region of the PWN. 

A review of Crab polarisation measurements in the hard X-ray band is given in~\citet{Chattopadhyay.2021}. Results are generally weakly constrained. Hitherto, the most precise measurement was provided by \pogo (20--160~keV), where a phase integrated PD=(20.9$\pm$5.0)\% and PA=(131.3$\pm$6.8)$^\circ$ was found~\citep{Chauvin.2017}.
In this paper, we report \XL results in the $\sim$19--64~keV range,
averaged over the full pulsar phase as well as for phase selections, including the off-pulse emission, the two pulsar peaks, and the intervening bridge region.
When combined with \I, our results provide a broadband view of the polarisation properties of Crab X-ray emission. 

\section{\textit{XL-Calibur}}

The \XL~\citep{Abarr.2021} balloon-borne polarimeter comprises a 12~m long optical bench (truss), pointed with arcsecond precision by the Wallops ArcSecond Pointer (WASP)~\citep{WASP2017}. A 45 cm diameter X-ray mirror~\citep{Tsunemi.2014} is mounted at one end of the truss, with a polarimeter/anticoincidence-shield assembly at the other end. 
The mirror focuses X-rays by Bragg reflection from multi-layer platinum and carbon foils.
With a field-of-view of $\sim$5~arcmin, the effective area is $\sim$350~cm$^2$ at 15~keV and $\sim$50~cm$^2$ at 60~keV. A sharp reduction at 78~keV results from the K absorption edge of platinum. The half-power diameter of the point-spread function (PSF) is $\sim$2~arcmin, reaching 7.3~mm diameter in the focal plane of the mirror. Prior to the flight, the shape of the PSF was measured at the SPring-8 synchrotron beam facility~\citep{Kamogawa.2022}. 

Focused X-rays pass through a tungsten collimator and impinge on a beryllium (Be) rod with diameter 12~mm and length 80~mm.
A fraction of incident X-rays ($\sim$85\% at 30~keV) will undergo Compton scattering from the Be rod into a circumadjacent assembly of 0.8~mm thick CdZnTe (CZT) detectors (each 20~mm~$\times$~20~mm area, 2.5~mm pixel pitch), arranged four-high around the rod in a square geometry. 
The energy resolution is $\sim$5.9~keV at 40~keV (full width at half maximum). 
To mitigate measurement background from atmospheric X-/$\gamma$-rays and neutrons, the polarimeter is housed in a 3--4~cm thick Bi$_{4}$Ge$_{3}$O$_{12}$ anticoincidence shield~\citep{Iyer.2023}, operated with a veto threshold of 50~keV. The polarimeter/shield assembly rotates around the viewing axis twice per minute to eliminate systematic errors arising from non-uniform CZT detector response.

The Klein-Nishina differential Compton-scattering cross-section depends on polarisation as
\begin{equation}
\frac{\mathrm{d}\sigma}{\mathrm{d}\Omega} \propto 2 \sin^2\theta \cos^2\phi,
\label{eq:kleinnishina}
\end{equation}
where $\theta$ denotes the polar scattering angle, and $\phi$ is the azimuthal angle between the photon scattering direction and the polarisation (electric field) vector of the incident X-ray. By reconstructing the distribution of azimuthal scattering angles, the linear polarisation of the incident beam can be estimated~\citep{DelMonte2022,Bernard2022}. The resulting modulation curve, $C(\phi)$, is a harmonic function with 180$^\circ$ periodicity:

\begin{equation}
C(\phi)=N\{1+\mu\cos[2(\phi-\zeta)]\},
\end{equation}
where $\mu$ and $\zeta$ are the amplitude and phase of the modulation curve, respectively, and $N$ is the mean. 
The modulation response to a 100\% polarised beam is $\mu_{100}$, so that $\mathrm{PD}=\mu/\mu_{100}$. 
Since X-rays preferentially scatter in a direction perpendicular to the orientation of the polarisation vector, $\mathrm{PA}=\zeta+90^\circ$. 

Reconstructed polarisation parameters are subject to a systematic error arising from movement of the focal point on the Be rod during observations, e.g. due to thermal or gravitational deformation of the truss~\citep{Aoyagi.2024}. For in-flight monitoring of the mirror alignment, an embedded back-looking camera (BLC) is used. Prior to flight, the direction of the mirror X-ray axis was determined at ISAS/JAXA and NASA Goddard Space Flight Center.  
At the launch site, the polarimeter-mirror alignment was confirmed using the BLC to image parallel beams of laser light focused by the mirror. X-rays that pass the Be rod impinge an imaging CZT. The profile of the X-ray beam reconstructed in this detector is used to determine the mean scattering position within the Be rod, which defines the interaction point assumed when calculating scattering angles.  

\section{Observations and data analysis}
\label{sec:obs}

\XL was launched on a 1.1~million-cubic-metre helium-filled balloon from the Esrange Space Center, Kiruna, Sweden
(68.89$^\circ$N, 21.11$^\circ$E), on July 9th, 2024, at 03:04~UT.  
Crab observations were conducted on July 11th, 12th and 13th, at a median altitude of $\sim$39.4~km (line-of-sight column density of $\sim$5.55~g/cm$^2$). The black-hole X-ray binary Cygnus X-1 was also observed\footnote{Publication in preparation.}. 
The flight followed a trajectory of approximately constant latitude before termination close to Kugluktuk, Nunavut, Canada, on July 14th, 2024, at 22:34 UT. 

During the commissioning phase, arcminute-level offsets were introduced to Crab pointing solutions to centre the focused X-ray beam on the Be rod, based on feedback from the imaging CZT. 
The resulting mean interaction-point offset, reconstructed within the perimeter of the 12~mm diameter Be rod, is (0.99, 0.21)~mm (zenith, starboard) when averaged across all observations, corresponding to a net offset of 17~arcsec. 

The pointing direction alternated between the Crab location (on-source observations) and background fields offset by 1$^{\circ}$ (off-source observations). 
Off-source pointings were conducted in a cross-pattern around the source, with excursions in azimuth or elevation.
An on/off ratio of 25~minutes/5~minutes was employed for the majority of observations. 

The X-ray interaction time in the polarimeter is determined with microsecond precision relative to GPS Universal Time using an on-board oscillator synchronised to a GPS pulse-per-second (PPS) signal. 
A phase-folded light-curve is obtained with a barycentre timing correction using the closest preceding ephemeris from the Jodrell Bank Observatory\footnote{\url{https://www.jb.man.ac.uk/pulsar/crab.html}}~\citep{Lyne.2014}: June 15th 2024. 
Phase selections are used when determining polarisation parameters for the pulsar. The off-pulse emission is subtracted from the pulsar-peak phase range, along with measurement background. 
As the GPS timing system experienced thermal-management problems on July 11th and 13th, phase-dependent studies are only possible for $\sim$63\% of the total signal integration-time.

Events comprising an above-threshold energy deposit (median threshold $\sim$15.4~keV, standard deviation $\sim$4.7~keV)
in a single CZT pixel and no coincident activity in the shield during a $\sim$3~$\mu$s window were used in the polarisation analysis. Noisy pixels were removed from the data based on measured scaler rates and inspection of pixel-by-pixel spectra, individually calibrated prior to flight using X-ray lines from a $^{152}$Eu source. To maximise the signal-to-background ratio, events with detected energy in the range 15--60~keV were kept. At lower energies, the signal rate diminishes due to atmospheric attenuation and pixel-by-pixel energy thresholds, while at higher energies the source flux drops below the background level. The selected range corresponds to $\sim$18.5--63.5~keV incident energy, after taking into account the average energy loss in the Be rod upon Compton scattering. 
The energy response was determined using Geant4~\citep{Geant4Ref} Monte-Carlo simulations~\citep{Aoyagi.2024}, where a Crab spectrum is generated and folded through the atmospheric attenuation and mirror response. 

\begin{figure}
\centering
\includegraphics[width=.9\columnwidth]{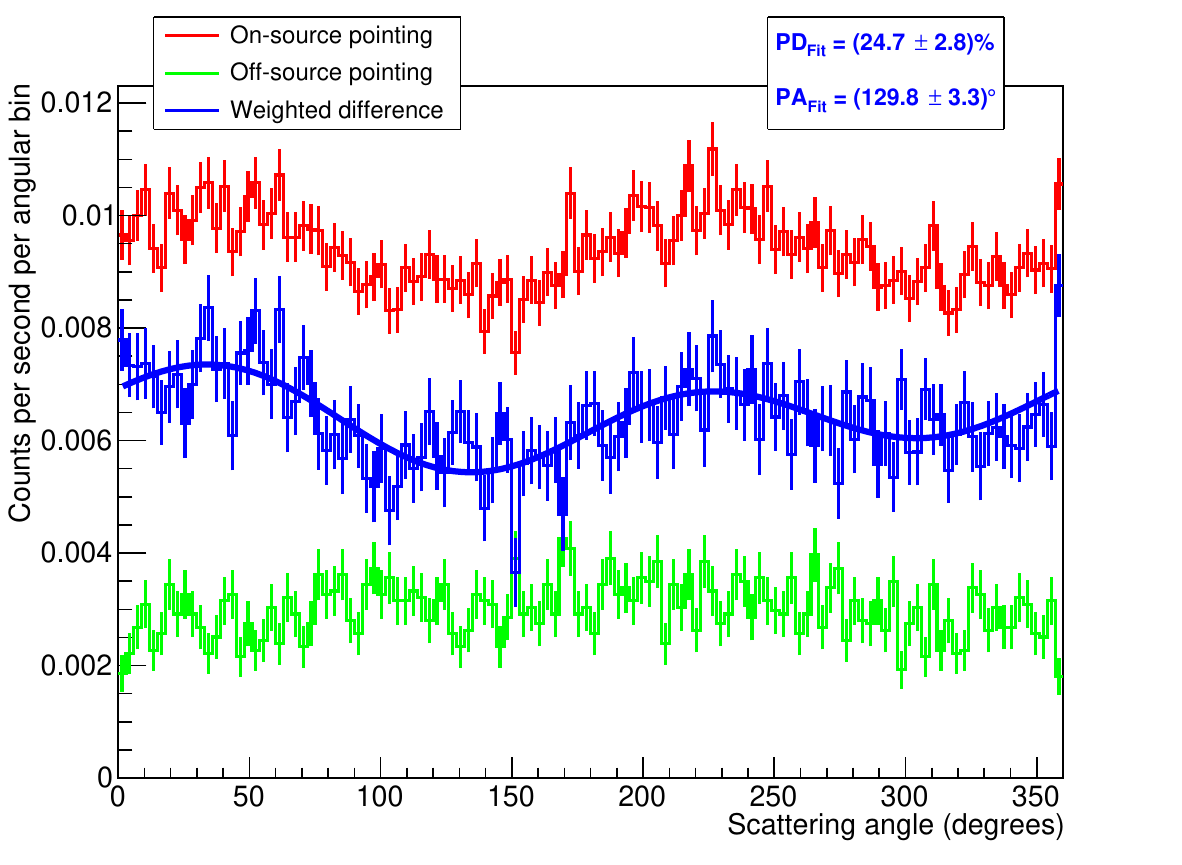}
\caption{\label{fig:modulationcurves}Scattering-angle distributions for the \XL Crab observation. The modulation in the on-source distribution (red) arises from the polarisation of the incident flux, as per Equation~(\ref{eq:kleinnishina}). The off-source distribution (green) is shown when calculated the same way as for on-source, i.e., relative to celestial north and using offset correction. Without offset correction and in the non-rotating truss frame, the off-source distribution becomes flat, indicating that the background impinging on the polarimeter is isotropic. Subtracting the exposure-weighted off-source distribution from the on-source result gives the signal excess (blue). The signal distribution is well-described by a fitted modulation curve with 180$^\circ$ periodicity, as well as a component with 360$^\circ$ periodicity (relative amplitude $\sim$1\%), arising due to the small residual offset of the focused X-ray beam from the centre of the Be rod~\citep{Aoyagi.2024}. In the absence of an offset correction, the relative amplitude of the 360$^\circ$ component becomes significantly higher ($\sim$10\% relative amplitude). The polarisation parameters determined from fitting the modulation curve are shown, where only fitting errors are quoted. Only the Stokes-parameter analysis described in Section~\ref{sec:obs} is used to produce the polarisation results presented in Section~\ref{sec:results}.} 
\end{figure}

For each valid detector event, the azimuthal scattering angle is first transformed from the rotating polarimeter frame to the non-rotating truss reference frame. Scattering angles are then converted to a coordinate system referenced to celestial north, using the truss orientation determined by the WASP system. The focal-point offset was accounted for by assuming scattering from the day-by-day-averaged interaction point, as measured with the imaging CZT. The endpoint was randomized within the 2.5-mm width of the hit wall-CZT pixel, as described in~\cite{Aoyagi.2024}. Resulting scattering-angle distributions are shown in Figure~\ref{fig:modulationcurves}.

For the polarisation analysis, selected events are used to form exposure-weighted sums of intensity ($I$)-normalised $Q$ and $U$ Stokes parameters ($\mathcal{Q}=Q/I$ and $\mathcal{U}=U/I$), following the procedure detailed in~\citet{Kislat.2015}. Reconstructed Stokes parameters $\mathcal{Q}_r$ and $\mathcal{U}_r$ additionally take the modulation response $\mu_{100}$ into account, e.g. $\mathcal{Q}_r={\mathcal{Q}}/{\mu_{100}}$.  
The additive nature of Stokes parameters makes it straight-forward to account for measurement backgrounds using on-source and exposure-weighted off-source observations, whereby polarisation parameters are defined as\footnote{Equation~(\ref{eq:PD}) is sometimes seen with an extra factor 2 (e.g.,~\cite{Kislat.2015,Chauvin.2017, Kiss2024}). Following~\cite{Baldini2022} and~\cite{Kislat2024}, we have here absorbed the factor two into the definition of the Stokes parameters themselves.}

\begin{equation}
    \mathrm{PD} = \frac{1}{\mu_{100}(I_\mathrm{on}-I_\mathrm{off})}\sqrt{(Q_\mathrm{on}-Q_\mathrm{off})^2+(U_\mathrm{on}-U_\mathrm{off})^2},
    \label{eq:PD}
\end{equation}

\begin{equation}
    \mathrm{PA}=\frac{1}{2}\arctan\left(\frac{U_\mathrm{on}-U_\mathrm{off}}{Q_\mathrm{on}-Q_\mathrm{off}}\right).
    \label{eq:PA}
\end{equation}

Since PD is a positive-definite quantity, the measurement may be subject to bias~\citep{Quinn.2012, Maier_2014, mikhalev2018pitfalls}. Equation~(\ref{eq:PD}) is therefore applicable only when PD~$\gtrsim$~MDP, i.e., when the measured polarisation degree exceeds the Minimum Detectable Polarisation~\citep{weisskopf2010understanding}. Defined at 99\% confidence level and using the formalism of~\citet{Kislat.2015}, 
\begin{equation}
    \label{eq:MDP}
    \mathrm{MDP}=\frac{4.29}{\mu_{100}R_s}\sqrt{\frac{R_b+f_{\mathrm{off}}R_s}{(1-f_{\mathrm{off}})f_{\mathrm{off}}T}}
\end{equation}
where $R_s$, $R_b$ and $T$ are the signal rate, background rate and observation time, respectively, and $f_{\mathrm{off}}$ is the fraction of time spent observing off-source (background) fields.
For a simulated 100\% polarised Crab beam~\citep{Aoyagi.2024}, $\mu_{100}=(42.4\pm$0.1)\%. 

We use a Bayesian framework for the polarisation analysis (see examples in ~\cite{Chauvin.2017, Abarr.2020p1n, Kiss2024}). The uninformative Jeffreys prior (uniform in Stokes Q and U) tends to overestimate the true PD~\citep{Maier_2014}. Following the reasoning in~\cite{Quinn.2012}, we instead assume a prior that is uniform in polar coordinates (PD, PA). Our posterior is evaluated in (PD, PA) space, as defined in~\cite{Chauvin.2017}. The maximum a posteriori (MAP) estimate is the mode (most probable value) of the two-dimensional Bayesian posterior, and corresponds to Eq.~(\ref{eq:PD}) and Eq.~(\ref{eq:PA}). While the MAP PD is a biased estimator, marginalising over PA prevents the point-estimate for from becoming biased~\citep{mikhalev2018pitfalls}. Our reported PD and PA values, as well as their credible regions (uncertainties) are derived by marginalization over the posterior. The Bayesian method provides asymmetric credible regions ensuring that the physical requirement PD~$\in$~[0, 1] is fulfilled.

\section{Results}
\label{sec:results}

\begin{figure*}
    \begin{center}
    \includegraphics[width=0.475\textwidth]{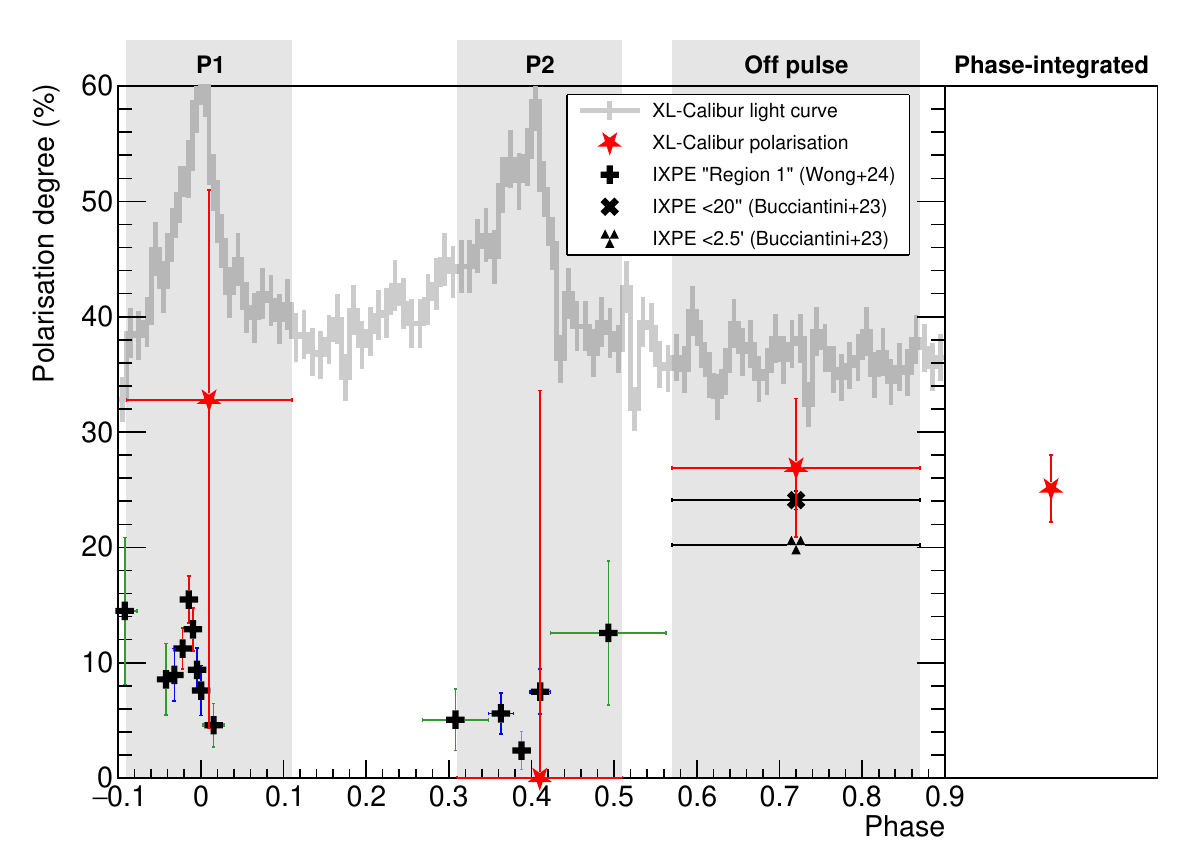}
    \includegraphics[width=0.475\textwidth]{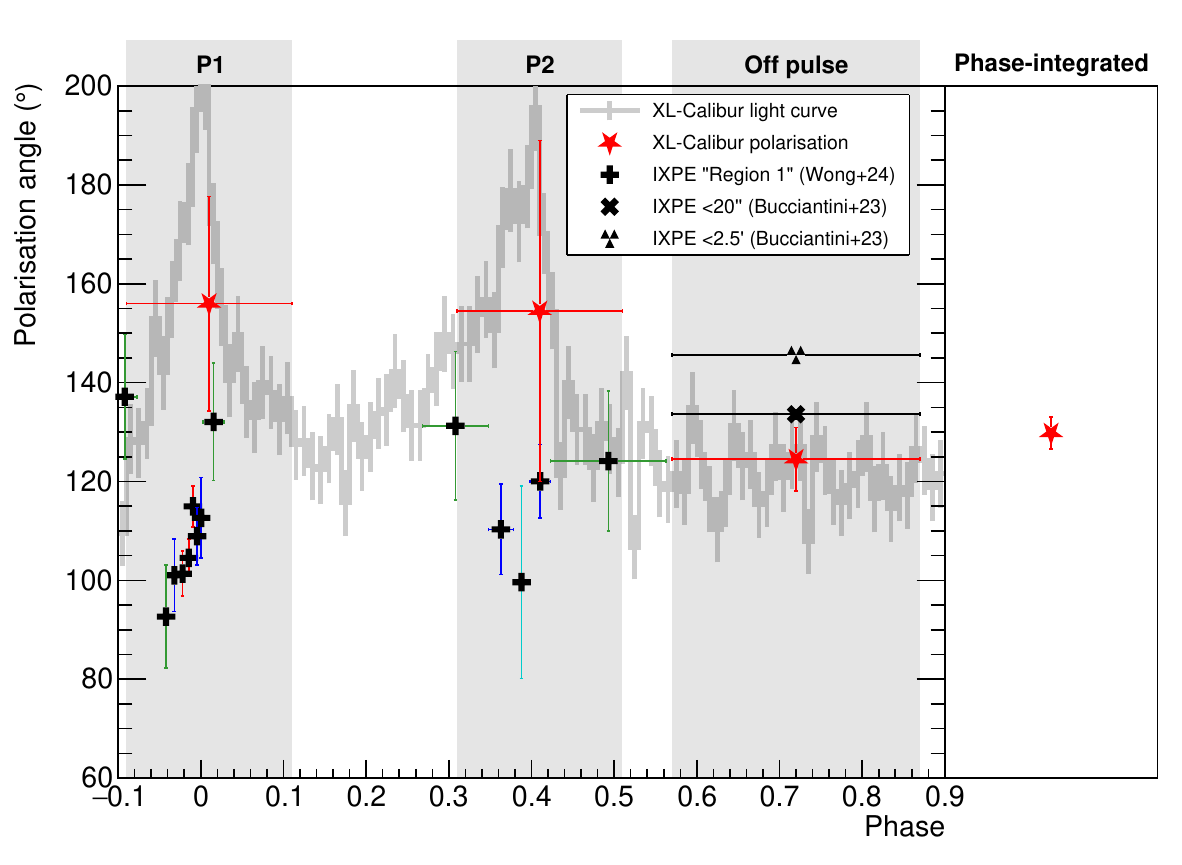}
    \caption{Marginalised polarisation degree and polarisation angle, superimposed on the phase-folded light-curve as measured by \XL. Shaded regions indicate relevant phase intervals from Table~\ref{table:statistics}. For the off-pulse PD and PA, exposure-weighted background has been subtracted, while for P1 and P2, the exposure-weighted off-pulse contribution has been subtracted together with the background, leaving only the pulsed contribution. Right-side panels show phase-integrated results. Corresponding \I results have been added from~\protect\cite{Bucciantini.2023} and~\protect\cite{wong2024analysiscrabxraypolarization}. For the latter, error-bar colours indicate different levels of significance ($>$5$\sigma$, $>$3$\sigma$, $>$1.9$\sigma$, $<$1.9$\sigma$ for red, blue, green and turquoise, respectively), and~\protect\cite{wong2024analysiscrabxraypolarization} have remarked that marginally significant measurements have significant PD$-$PA covariance, whereby reported error bars are not fully representative.}
    \label{fig:LC}
    \end{center}    
\end{figure*}

The total on-source observation time was 49.7~ks, interspersed with 17.1~ks of off-source observations. On-source rates vary depending on the Crab elevation and observing altitude, with an average of $\sim$1.1~counts per second observed, while the background rate of $\sim$0.36~counts per second remained constant during observations. The resulting mean signal-to-background ratio of 3 is $\sim$20 times higher than \pogo~\citep{Chauvin.2017}. 
Observation statistics are summarised in Table~\ref{table:statistics}. The \XL light-curve is defined with the measured peak of P1 at phase 0. We then adopt the same phase intervals as \I for the off-pulse/bridge~\citep{Bucciantini.2023} and P1/P2~\citep{wong2024analysiscrabxraypolarization}.
Marginalised polarisation parameters are shown superimposed on the measured pulsar light-curve in Figure~\ref{fig:LC}.
The pulsed fraction is (14.6$\pm$1.0)\%, matching~\citet{Eckert.2010}. 
Results are summarised in Table~\ref{table:results}. 

\begin{table*}
\centering
\caption{\label{table:statistics}Summary of observation statistics. The light-curve is defined with the peak of P1 as measured by \XL at phase 0. For off-source (background) observations, $\sim$57\% of the integration time was without PPS timestamps. Since off-source fields are not correlated to the pulsar phase, all Crab off-source pointings can still be used in the analysis, which improves the precision of the background subtraction. As an independent confirmation, full and reduced off-source datasets have been inspected separately, and agree within the statistical precision of the observations.
}
\begin{tabular}{|c|c|c|c|c|}
\hline
{\bf Observation} & {\bf Phase range} & {\bf Detected events} & {\bf Exposure (s)} & {\bf Remarks}\\
\hline
\hline
Full Crab on-source dataset & N/A & 56054 & 49716 & No requirement on PPS timing\\
\hline
Crab on-source with PPS timing & $-0.1$ to $0.9$ & 34523 & 31273 & Subset of preceding entry\\
\hline
Off-pulse region & $0.57$ to $0.87$ & 9280 & 9382 & Interval from~\cite{Bucciantini.2023}\\
\hline
Bridge region & $0.07$ to $0.27$ & 6690 & 6255 & Interval from~\cite{Bucciantini.2023}\\
\hline
P1 ("main pulse") & $-0.09$ to $0.11$ & 7686 & 6255 & Interval from~\cite{wong2024analysiscrabxraypolarization}\\
\hline
P2 ("inter-pulse") & $0.31$ to $0.51$ & 7673 & 6255 & Interval from~\cite{wong2024analysiscrabxraypolarization}\\
\hline
Crab off-source (background) dataset & N/A & 6162 & 17144 & No requirement on PPS timing\\
\hline
\end{tabular}
\end{table*}

\begin{table*}

\centering
\caption{\label{table:results}Summary of results. The maximum a posteriori (MAP) estimate corresponds to Eq.~(\ref{eq:PD}) and Eq.~(\ref{eq:PA}). Tabulated PD and PA values result from marginalising over the Bayesian posterior, and uncertainties stated are credible regions corresponding to 1$\sigma$ Gaussian probability content. The MDP is calculated using Eq.~(\ref{eq:MDP}). Background subtraction is done using the phase-integrated background, without requiring GPS-timing, except for P1 and P2, where the off-pulse phase-region from on-source pointings is used as background, leaving only the pulsed component. The bridge result is not off-pulse subtracted.}
\begin{tabular}{|l|c|c|c|c|c|c|c|}
\hline
{\bf Phase} & {\bf $\mathcal{Q}_r$} & {\bf $\mathcal{U}_r$} & {\bf MAP (\%, $^{\circ}$)} &{\bf PD (\%)} & {\bf PA ($^{\circ}$)} & {\bf MDP (\%)}\\
\hline
\hline
Full Crab on-source dataset & -0.045 $\pm$ 0.029 & -0.249 $\pm$ 0.029 & (25.3, 129.8) & 25.1 $\pm$ 2.9  & 129.8 $\pm$ 3.2 & 8.7\\
\hline
Crab on-source with PPS timing & -0.065 $\pm$ 0.034 & -0.277 $\pm$ 0.033 & (28.5, 128.4) & 28.3 $\pm$ 3.4 & 128.4 $\pm$ 3.4 & 10\\
\hline
Off-pulse region ("nebula") & -0.099 $\pm$ 0.060 & -0.257 $\pm$ 0.059 & (27.5, 124.5) & 26.9 $\pm$ 6.0 & 124.5 $\pm$ 6.4 & 18\\
\hline
Bridge region & -0.006 $\pm$ 0.065 & -0.267 $\pm$ 0.065 & (26.7, 134.4) & 25.9 $\pm$ 6.6 & 134.4 $\pm$ 7.3 & 20\\
\hline
P1 ("main pulse") & 0.290 $\pm$ 0.241 & -0.322 $\pm$ 0.241 & (43.3, 156.0) & 32.8$^{+18.2}_{-28.5}$ & 156.0 $\pm$ 21.7 & 73\\
\hline
P2 ("inter-pulse") & 0.178 $\pm$ 0.243 & -0.219 $\pm$ 0.243 & (28.2, 154.5) & 0.0$^{+33.6}_{-0.0}$ & 154.5 $\pm$ 34.5 & 74\\
\hline
\end{tabular}
\end{table*}

The PD is constrained at $\sim$1$\sigma$ level for P1. 
For P2, the marginalised PD peaks at 0\%.
Upper limits at 99\% Confidence Level are 90.1\% and 78.5\%, respectively.
As shown in Figure~\ref{fig:LC}, our observations are in line with \I, where (in the lower 2--8~keV band) the overall PD level for P1 appears to be higher than P2~\citep{wong2024analysiscrabxraypolarization}.

{\it XL-Calibur} measures the polarisation of the phase-averaged and off-pulse emissions with high statistical significances of 8.6$\sigma$ and 4.5$\sigma$, respectively. The two results are compatible within the statistical accuracy of the measurements, as expected if the overall pulsar contribution is only weakly polarised.
We do not find evidence for a change of the PD and PA between 15--31~keV and 31--60~keV (Table~\ref{table:energybinning}). 
Our phase-averaged results for PD and PA are compatible with PoGO+ observations conducted in 2016~\citep{Chauvin.2017}. This may indicate that the overall magnetic configuration of the inner emission region was similar when the two observations were made. Between these measurements, an excursion in soft X-ray on-pulse PD was reported by {\it PolarLight} (3.0--4.5~keV)~\citep{feng2020re, Long_2021}.

\begin{table}
\centering
\caption{Phase-integrated polarisation parameters when sub-dividing into two intervals of detected energy. The number of signal events (after background subtraction) is denoted N$_{\mathrm{signal}}$.}
\label{table:energybinning}
\begin{tabular}{|c|c|c|c|c|}
\hline
{\bf Energy (keV)} & {\bf N$_{\mathrm{signal}}$} & {\bf PD (\%)} & {\bf PA ($^{\circ}$)} & \bf MDP (\%)\\
\hline
\hline
$15-31$ & 19405 & 24.2 $\pm$ 3.9 & 128.3 $\pm$ 4.5 &11.6\\
\hline
$31-60$ & 18779 & 28.0 $\pm$ 4.3 & 128.3 $\pm$ 4.3 & 13.0\\
\hline
\end{tabular}
\end{table}

\section{Discussion}
\label{section:disc}
By combining a large-area X-ray mirror with a compact and well-shielded polarimeter, \XL measures the polarisation of the $\sim$19--64~keV X-ray emission of the Crab pulsar and PWN with unprecedented sensitivity. 
The systematic error on our measurements stems from (i) the focal-point offset correction, (ii) the uncertainty on $\mu_{100}$, and (iii) background subtraction. For (i), with no offset correction applied, the PD only changes by $\sim$0.7 percentage points. 
Since the uncertainty on $\mu_{100}$ from simulations is negligible, (ii) is dictated by knowledge of the PSF. For two independent PSF measurements,
the simulated $\mu_{100}$ varies by only 0.2 percentage points.
The systematic error arising from background subtraction (iii) is minimal since signal and background observations are interspersed. We therefore expect the overall systematic error to be significantly smaller than the statistical error. 

Our observations indicate a low level of polarisation when integrating across the pulsar peaks. As noted in~\citet{Bucciantini.2023}, this contrasts with many inner magnetospheric models, where high-energy emission arises from accelerated particles emitting synchrotron emission within plasma-starved gaps in the magnetosphere. 
In the striped-wind picture of \cite{Petri2013}, emission outside the light cylinder
(LC) generates strong reductions in the PD at and following each pulse peak. The
plasma simulations of \cite{Cerutti.2016pz1} at the current sheet inside
the LC display a modest anti-correlation between PD and intensity, and can
have large variations in PA. 
For a force-free magnetic geometry, \citet{Harding.2017feh} demonstrate that
synchrotron radiation from regions just outside the LC exhibit PDs at the measured levels, however the accompanying PA variations at various phases are
strong. 
Higher-significance PD measurements within the pulsar peaks are required to discriminate between these various scenarios. 

For \XL this is not a straight-forward proposition.  
Even when assuming fully efficient observations with purely on-source pointing during a week-long 
flight\footnote{This is currently the maximum observation time for high-latitude balloon missions observing the northern sky.}, MDP$\sim$20\% is expected when integrating across P1 or P2. To accurately track the phase-dependence of polarisation parameters, the MDP must be improved by an order of magnitude. Further reducing the already low measurement background is not a feasible approach. The signal rate can be enhanced by increasing the effective area of the X-ray mirror. To be practical, this would require multiple mirror assemblies -- a significant engineering challenge, since \XL is already close to the payload mass-limit for the current balloon. Work is in progress to allow phase-folding of data from July 11th and 13th when GPS timing was unavailable. While this may lead to a modest reduction in the PD uncertainty for the pulsar peaks, our overall conclusions will not be affected.

\begin{figure*}
    \begin{center}
    \includegraphics[width=.65\textwidth]{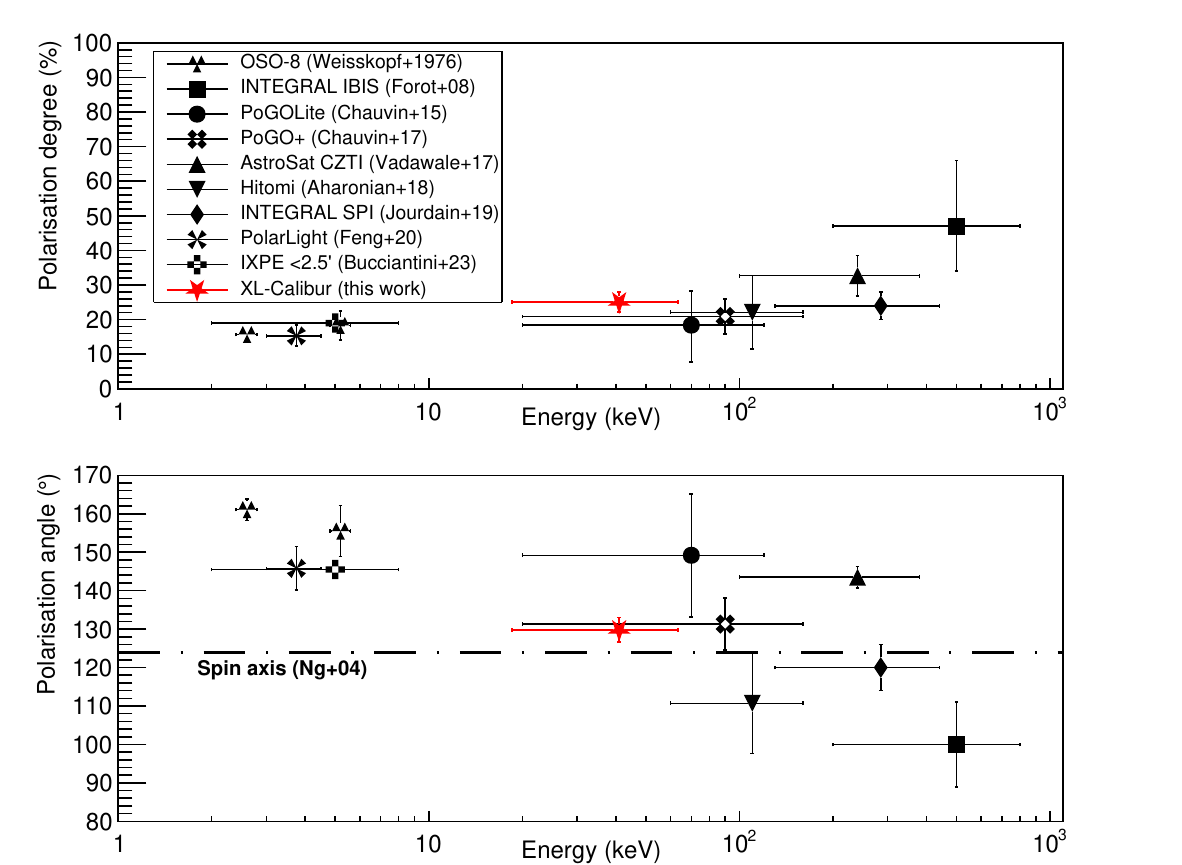}
    \caption{Comparison of phase-integrated Crab results, from soft X-rays to gamma-rays, for polarisation degree (top) and angle (bottom). Missions are listed in chronological order, with data points centred on the corresponding energy range. References:~\protect\cite{weisskopf1976}, \protect\cite{forot2008polarization},~\protect\cite{chauvin2015observation}, \protect\cite{Chauvin.2017}, \protect\cite{Vadawale.2017}, \protect\cite{hitomi2018detection}, \protect\cite{jourdain20192003}, \protect\cite{feng2020re}, \protect\cite{Bucciantini.2023}. The spin axis from~\protect\cite{Ng_2004} has also been indicated.}
    \label{fig:comp}
    \end{center}    
\end{figure*}

\begin{figure}
\centering
\includegraphics[width=.975\columnwidth]{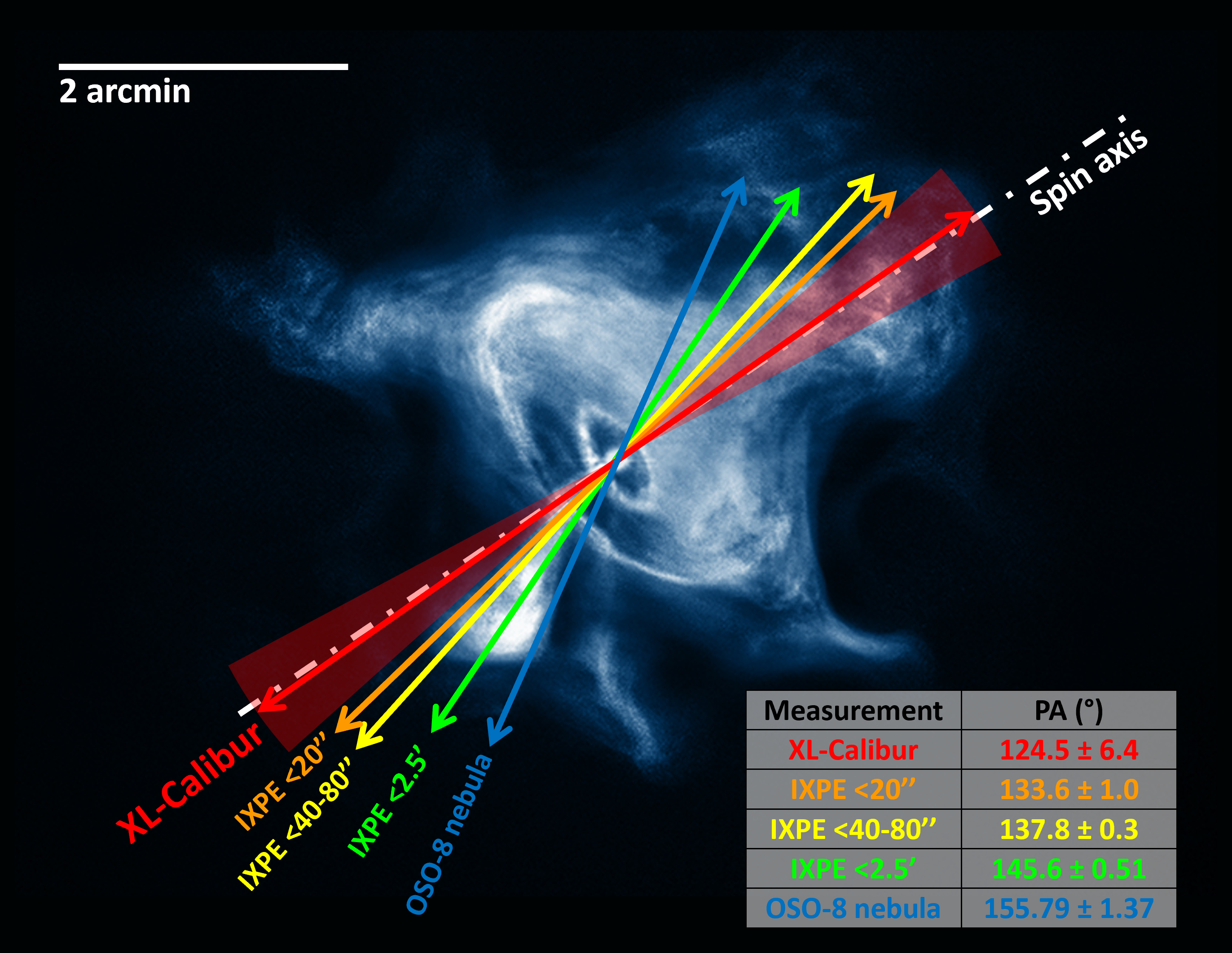}
\caption{\label{fig:nebulaandpa}Off-pulse polarisation results from {\it OSO-8}~\protect\cite{weisskopf1978precision} (nebula only, 2.6~keV and 5.2~keV combined, in blue), \I~\protect\cite{Bucciantini.2023} ($<$2.5~arcmin in green, $<$20~arcsec in orange), \protect\cite{wong2024analysiscrabxraypolarization} ("Region 1" in yellow) and \XL (red), superimposed on a \C image of the inner region of the Crab nebula (approximately 200$\times$200~arcsec). Arrow lengths are proportional to PD. For \XL, the $1\sigma$ error on PA has been indicated. Corresponding errors for the remaining measurements are small compared to the arrow widths. The pulsar spin-axis from~\protect\cite{Ng_2004} is shown in white. Image credit: \C X-ray: NASA/CXC/SAO.
}
\end{figure}

A comparison to previous phase-integrated X-/$\gamma$-ray measurements is shown in Figure~\ref{fig:comp}.
Our phase-averaged PD=(25.1$\pm$2.9)\% and PA=(129.8$\pm$3.2)$^{\circ}$ 
differ from the PD=(19.0$\pm$0.19)\% and PA=(145.5$\pm$0.3)$^{\circ}$ measured by \I within 2.5~arcmin of the pulsar~\citep{Bucciantini.2023}.
While the regions viewed by both missions are comparable, the higher energy band of \XL ($\sim$19--64~keV) compared to \I (2--8~keV) means that the emission originates closer to the central toroidal region of the PWN, since the synchrotron radiative lifetime for outflowing electrons is inversely proportional to energy.
This is shown by \Nu, where 3--8~keV observations reveal inner nebula emission from a larger region ($\sim$75$\times$40~arcsec) than at 20--80~keV 
($\sim$60$\times$25~arcsec)~\citep{Madsen.2015s4a}. \Hit HXT, 15--70~keV, confirms this trend~\citep{Morii.2024r1}. 
We note that \I off-pulse measurements for (i) an elliptical region extending $\sim$80~arcsec ($\sim$40~arcsec) along the toroidal (jet) axis~\citep{wong2024analysiscrabxraypolarization}, referred to as {"Region~1"}, and (ii) within 20~arcsec of the pulsar~\citep{Bucciantini.2023}, are more compatible with \XL off-pulse results. This is illustrated in Figure~\ref{fig:nebulaandpa}, showing that the \XL PA is aligned with the pulsar spin axis, indicating that the harder X-rays measured by \XL are dominated by the central region of the PWN. For {\it OSO-8} measurements encompassing the full nebula~\citep{weisskopf1978precision}, the PA (155.79$\pm$1.37)$^{\circ}$ at 2.6~keV and 5.2~keV combined lies even further away from the \XL PA.

As discussed by \citet{Madsen.2015s4a}, the imaging-spectroscopic \Nu observations of the Crab PWN are compatible with downstream leptons propagating sub-relativistically mostly through advection~\citep{Kennel.1984,Kennel.1984cmc}
rather than through diffusion~\citep{Gratton.1972,Wilson1972}.
Detailed 3D modeling of the combined \C, \I, \Nu and \XL data with magnetohydrodynamics (e.g.~\citet{Porth2013}) and Particle in Cell (e.g.~\citet{Cerutti.2021}) codes is outside of the scope of this paper but could give new constraints on the structure of PWN, and their relativistic shocks, for the Crab and more generally.

\section*{Acknowledgements}

\XL is funded in the US by the NASA APRA (Astrophysics Research and Analysis) program through grants 80NSSC20K0329 and 80NSSC24K0205. FK acknowledges funding from NASA ADAP award 80NSSC24K0636 and NASA IXPE GO Cycle 1 award 80NSSC24K1762. 
The Japanese Society for the Promotion of Science (JSPS) has supported this work through KAKENHI Grant Numbers 19H01908, 19H05609, 20H00175 (HM), 20H00178 (HM), 21K13946 (YU), 22H01277 (YM), 23H00117 and 23H00128 (HM).
KTH authors are supported by the Swedish National Space Agency (2022-00178). MP also acknowledges funding from the Swedish Research Council (2021-05128). 
We thank the NASA-WASP team for campaign support and pointing operations. 
We are very grateful to colleagues at NASA-CSBF and Esrange Space Center who provided balloon launch, operations and recovery services. 
H.\,Tsunemi and A.\,Furuzawa are thanked for their role in the construction of the mirror.
T.\,Enoto, Y.\,Fukazawa, S.\,Gunji, T.\,Mizuno and Y.\,Saito are thanked for advice and discussions. 
M.\,Aoyagi, K.\,Ishiwata, W.\,Kamogawa, H.\,Matake, N.\,Sakamoto are thanked for work on previous mirror calibration activities.
J.\,Wong is thanked for providing tabulated \I results for reference.
No AI was used in the data analysis presented here, nor for the writing of this paper.

\section*{Data Availability}

The \XL data underlying this article will be made available via the NASA HEASARC data archive, at 
\url{https://heasarc.gsfc.nasa.gov/docs/xlcalibur/}.



\bibliographystyle{mnras}
\bibliography{XL_Crab_MNRAS} 








\bsp	
\label{lastpage}
\end{document}